\documentstyle[prd,aps,epsfig]{revtex}
\bibstyle{unsrt}

\newcommand{\eqref}[1]{(\ref{#1})}
\newcommand{\dd}{{\rm d}}
\newcommand{\half}{\frac12}
\newcommand\mean[1]{{\big<#1\big>}}

\begin{document}
 
\preprint{LAUR 99-891}  
\twocolumn[\hsize\textwidth\columnwidth\hsize\csname 
@twocolumnfalse\endcsname
\draft
 
\title{Controlling One-Dimensional Langevin Dynamics on the
Lattice} 
\author{Lu\'{\i}s M. A. Bettencourt,$^{1,*}$ 
Salman Habib,$^{2,\dagger}$ and Grant Lythe$^{3,\#}$}
\address{$^1$T-6/T-11, Theoretical Division, MS B288, 
Los Alamos National Laboratory, Los Alamos, New Mexico 87545} 
\address{$^2$T-8, Theoretical Division, MS B285, 
Los Alamos National Laboratory, Los Alamos, New Mexico 87545}
\address{$^3$CNLS, Theoretical Division, MS B258, 
Los Alamos National Laboratory, Los Alamos, New Mexico 87545}
\date{\today}
\maketitle
\begin{abstract}
Stochastic evolutions of classical field theories have recently become
popular in the study of problems such as determination of the rates of
topological transitions and the statistical mechanics of nonlinear
coherent structures. To obtain high precision results from numerical
calculations, a careful accounting of spacetime discreteness effects
is essential, as well as the development of schemes to systematically
improve convergence to the continuum. With a kink-bearing $\phi^4$
field theory as the application arena, we present such an analysis for
a 1+1-dimensional Langevin system. Analytical predictions and results
from high resolution numerical solutions are found to be in excellent
agreement. 
\end{abstract}

\pacs{PACS Numbers : 11.15.Pg, 11.30.Qc, 05.70.Ln, 98.80.Cq 02.50-r
\hfill   LAUR 99-891} 

\vskip2pc]

\section{Introduction}

In recent years there has been growing interest in extracting 
non-perturbative quantum dynamical information such as 
topological transition rates from numerical Langevin and 
Monte Carlo solutions of classical field
theories at finite temperature \cite{interest}.  At the next level of
sophistication, several attempts have been made at developing schemes
that treat low-lying modes classically and high frequency modes
quantum mechanically \cite{match}.  Moreover, the equilibrium and
nonequilibrium classical statistical mechanics of nonlinear coherent
structures such as kinks has historically received much attention
\cite{kinks} in the condensed matter literature. Until fairly
recently, computer memory and performance restrictions were
sufficiently severe that Langevin evolutions could only be carried out
at fairly low levels of accuracy and resolution. However, present-day
supercomputers have overcome this problem, at least for low
dimensional problems, and one can well contemplate systematically
studying, understanding, and improving the accuracy of stochastic
evolutions. In this paper we present just such a study applied to
1+1-dimensional Langevin systems.

Our focus will be on lattice errors for quantities computed at thermal
equilibrium. In calculations of this type, a stochastic partial
differential equation (SPDE) augmented with a fluctuation-dissipation
relation is solved as an initial value problem using finite
differences. Because of the fluctuation-dissipation relation, the
system is eventually driven to thermal equilibrium and at late times
one may measure values of thermodynamic quantities as well as time and
space dependent correlation functions. These quantities can depend on
the lattice spacing, \(\Delta x\), on the total system size, on the
discretization used for spatial operators, and on the timestepping
algorithm used to solve the resulting set of coupled stochastic
ordinary differential equations.  In one space dimension a fairly
complete description can be given since the question of lattice
effects is one of convergence properties of SPDEs rather than of
renormalization. 

The configurational part of the partition function of a classical
field theory in one space dimension is free from divergences. In
particular, quantities such as kink densities, measured from finite
difference solutions of the corresponding SPDEs, converge to a
well-defined limit as the lattice spacing is reduced towards zero. The
question of exactly how the convergence scales with $\Delta x$ is
still a matter of practical importance: numerical solutions are
limited by the available computing power and memory to a finite range
of values of $\Delta x$. While finite volume effects can be important
in small lattices they are not important if the lattice size is much
larger than the longest correlation length. We will assume that this
is always the case in the considerations below.

A complete constructive procedure for determining the spatial lattice
error, and possibly eliminating it to some order in $\Delta x$,
exists.  The method proceeds as follows.  In equilibrium, the
probability of a given set of configurations can be calculated from
the static solution of the Fokker-Planck equation corresponding to the
particular spatial discretization and time-stepping algorithm applied
to the SPDE of interest. With time-stepping errors tuned to be
sub-dominant, the transfer integral \cite{ti} corresponding to the
lattice Hamiltonian can be evaluated to some given order in \(\Delta
x\).  Correlation functions and thermodynamic quantities, which can
all be extracted from the transfer integral, explicitly exhibit
lattice dependences allowing the convergence to the continuum to be
read off directly. We describe this procedure in more detail below.

Discreteness effects have been considered before in the context of
kink dynamics \cite{discretekink}. Trullinger and Sasaki have already
obtained the lowest-order discreteness corrections to the
Schr\"odinger equation that emerges from the transfer integral
approach \cite{TS}. They found that the lowest order correction is of
order \(\Delta x^2\) and, in first-order perturbation theory, is
equivalent to a corrected effective potential. As we show below, the
latter result can be adapted not only to compute the order of the
lattice errors but also to introduce a local counterterm in the
stochastic evolution equations that can drastically improve the
convergence to the continuum. Our results from high resolution
numerical solutions are in excellent agreement with the theoretical
predictions.

The class of problems considered here are 1+1-dimensional classical
field theories defined by the Hamiltonian:   
\begin{equation}
H=\int \dd x \left[{1\over 2}\pi^2 +{1\over
2}\left({\partial\Phi\over\partial x}\right)^2+ V(\Phi)\right]~.
\label{Hamiltonian}
\end{equation}
The corresponding continuum SPDE
\begin{equation}
{\partial^2\over\partial t^2}\Phi(x)={\partial^2\over\partial
x^2}\Phi(x)-\eta{\partial\over\partial
t}\Phi(x)-{\delta\over\delta \Phi}V(\Phi)+F(x,t) 
\label{contspde}
\end{equation}
is second order in time, where with $\beta=1/kT$, the noise and
damping obey a fluctuation-dissipation relation:
\begin{equation}
\langle F(x,t)F(y,s)\rangle=2\eta\beta^{-1}\delta(t-s)\delta(x-y)~.
\label{contfdr}
\end{equation}

In this paper we will adopt the example of the double-well $\Phi^4$ 
theory: $V(\Phi)=-(m^2/2)\Phi^2+(g^2 /4)\Phi^4$.  We shall work in a
dimensionless form of the theory given by the transformations:
$\phi=\Phi /a$, $\bar{x}=mx$, and $\bar{t}=mt$, where
$a^2=m^2/g^2$. Under these transformations, the original Hamiltonian
becomes $\bar{H}=H/(ma^2)$ where $\bar{H}$ is of the same form as the
original Hamiltonian $H$, except that the potential
$V(\phi)=-(1/2)\phi^2 + (1/4)\phi^4.$

This theory admits the well-known (anti-)kink solutions which, at zero
temperature, are exact solutions of the static field equations
connecting $\phi=-1$ at $x=(+)-\infty$ to $\phi=+1$ at $x=(-)+\infty$.
In thermal equilibrium, the balance between noise and damping is
manifested in the balance of nucleation and annihilation of
kink-antikink pairs \cite{handl}. At low temperature, WKB techniques
applied to the transfer integral \cite{KS,CKBT} yield the following
approximation for the density of kinks:
\begin{equation}
\rho_k \propto (E_k/kT)^{1/2}\exp(-E_k/kT),
\label{ndens}
\end{equation}
where $E_k=\sqrt{8/9}$, the energy of an isolated kink.  Supporting
numerical evidence exists \cite{fash}, but precise results have been
difficult to obtain until recently due to the large amount of
computing time needed at temperatures low enough to clearly
distinguish kinks. The best results obtained so far are for a special
double-well potential where exact theoretical computations can also be
carried out. In this case, it has been shown that the theoretical and
numerical results agree within statistical bounds set by the finite
volume of the simulations \cite{qes}.

The classical partition function for a $\phi^4$ theory, in any spatial
dimension $2 \leq D<4$, is super-renormalizable, i.e., there are a finite
number of perturbative diagrams that are divergent in the continuum,
but can be appropriately subtracted by the inclusion in the theory of
a finite number of suitable counterterms. The situation is different
for $D=1$: the continuum partition function is finite and no
renormalization is necessary.

An alternative approach to the one described here has been suggested
by Gleiser and M\"uller \cite{GM} who have proposed a perturbative
counterterm for use in 1+1-dimensional Langevin equations.  A
weakness of the latter proposal is that it relies on an approximation
to the free energy; in many situations the latter is a poor indicator
of the true dynamics of field theories \cite{noneq}. Moreover, their
counterterm is based on an approximate effective potential calculated
by perturbing about a uniform state. We will show below, with both
analytic and numerical results, the inadequacy of perturbative
counterterms in dealing with the convergence to the continuum.

The paper is organized as follows. In Section II we consider the
evolution of the probability density of the discretized SPDE. We use
the method of Horowitz \cite{Horowitz} to examine the effect of time
discretization on the equilibrium density. The transfer integral is
introduced in Section III. We perform calculations at finite \(\Delta
x\) and show that the leading order corrections to the continuum of
observable quantities are proportional to \(\Delta x^2\).  Examination
of the form of the Schr\"odinger equation at finite \(\Delta x\)
reveals a natural choice for a local counterterm with which to improve
the convergence properties of discretized Langevin equations. The
alternative one loop approach of Gleiser and M\"uller is examined in
Section IV. Numerical results are presented in Section V. In Section
VI we end with a discussion of our results.

\section{The Discrete Time Fokker-Planck equation}

Our first step in determining the (equilibrium canonical) distribution
to which a given Langevin dynamics converges for long times is to
derive the corresponding Fokker-Planck equation. This can be done on
the lattice as well as in the continuum.

On the lattice, an SPDE is solved by updating
\(2N\) quantities \(\{\phi_i(t),\pi_i(t)\}\) where \(i=1,\ldots,N\).
We take the lattice Hamiltonian in one space dimension, $H_{\rm lat}$, to be
\begin{equation}
H_{\rm lat} = \Delta x \sum_{i=0}^N  \left[ 
{1 \over 2} \pi^2_i +S(\phi_i) \right],
\end{equation}
with 
\begin{eqnarray}
S(\phi_i)&=& { 1 \over 2}  
{ \left( \phi_{i+1} -\phi_i \right)^2 \over \Delta x^2} + V(\phi_i), 
\\   
V(\phi_i) &=& -{ 1 \over 2 } \phi_i^2 + {1 \over 4} \phi_i^4.
\end{eqnarray} 
The corresponding Fokker-Planck equation for the $2N$ variables has a
static solution that can in principle be attained at late times in a
Langevin simulation (in the sense of ensemble averages over individual
simulations).

In practice the time as well as the space discretization of a Langevin
equation leads to errors.  The simplest stochastic timestepping is of
the Euler type and can be written as:
\begin{eqnarray}
\pi_i({t+\Delta t}) &&= \pi_i(t) - \Delta t 
\left[ \eta \pi_i(t)-{\partial H_{\rm lat} \over \partial \phi_i(t)}
\right]  
+ \xi_i(t), \nonumber \\
\phi_i({t+\Delta t}) &&= \phi_i(t) + \Delta t \pi_i({t}).
\label{e1} \end{eqnarray} 
We have chosen the case of additive Gaussian white noise, related to
the damping $\eta$ by the (suitably discretized)
fluctuation-dissipation relation:
\begin{eqnarray} 
\mean{\xi_t(t)} =0, \quad 
\mean{\xi_i(t)\xi_{j}(t')} = {2 \eta \over \beta}{1\over\Delta t\Delta
x}\delta_{ij}\delta_{tt'}.   
\label{e2} \end{eqnarray} 

 In order to understand the effect of time discretization, it is
possible write a discrete time Fokker-Planck equation, describing the
evolution of the probability density functional associated with
\eqref{e1}-\eqref{e2} \cite{Horowitz}:
\begin{eqnarray}
&& P[\{\pi,\phi\},t+\Delta t]
= \exp\left(-\Delta t   {\partial \over\partial \phi_i}  
{\partial H _{\rm lat}\over \partial \pi_i} \right)\times \label{e3} \\
&& \exp\left[\Delta t  {\partial \over \partial \pi_i} 
\left(\eta {\partial H_{\rm lat} \over \partial  \pi_i} + 
{\partial H_{\rm lat} \over \partial \phi_i}\right) + \Delta t
{\eta \over \beta} {\partial^2 \over \partial \pi^2_i} \right]
P[\{\pi,\phi\},t], 
\nonumber
\end{eqnarray}  
where summation over repeated indices is implied. For simplicity
this will be assumed in what follows and indices dropped.
The discrete time equation \eqref{e3} can be written in the form
\begin{eqnarray}
P[\{\pi,\phi\},t+\Delta t]
= e^{-\Delta t  H_{\rm FP}} 
P[\{\pi,\phi\},t].
\label{e4}
\end{eqnarray}
The operators in the two exponents in \eqref{e3} are non-commuting. To
reduce \eqref{e3} to the form \eqref{e4} we use the
Campbell-Baker-Hausdorff theorem: given the operators $A$ and $B$,
there is (formally) an operator $C$ such that $e^A e^B=e^C$, with
\begin{eqnarray}
C=&& A+B+ {1 \over 2} [A,B] + {1\over 12} [A,[A,B]]  \nonumber \\
&+&{1\over 12} [B,[B,A]]+\ldots 
\label{e5}
\end{eqnarray}
Expanding to first order in $\Delta t$, we have 
\begin{eqnarray}
H_{\rm FP} &=& {\eta \over \beta} 
{\partial^2 \over \partial \pi^2} 
-{\partial H_{\rm lat} \over \partial \pi}  
{\partial \over \partial \phi} + 
{\partial \over \partial \pi} \left( \eta  
{\partial H_{\rm lat} \over \partial \pi}
+ {\partial H_{\rm lat} \over \partial \phi} \right) \nonumber \\ 
&& + {1 \over 2} \Delta t   
\left[ \eta {\partial H_{\rm lat} \over \partial \pi} 
+ {\eta \over \beta} {\partial \over \partial \pi}
+ {\partial H_{\rm lat} \over \partial \phi} \right] 
{\partial \over \partial \phi} \label{e6} \\
&& -{1\over 2} \Delta t~{\partial H_{\rm lat} \over \partial \pi}
{\partial^2 H_{\rm lat} \over \partial
\phi^2} {\partial \over \partial \pi} + {\cal O}({\Delta t^2}).
\nonumber
\end{eqnarray}
Notice that each factor of $H_{\rm lat}$ introduces a power
of $\Delta x$. 

The solution of $H_{\rm FP} P[\{\phi,\pi\}]=0$ is the canonical
distribution approached by the discretized system at late times. Its
form can be computed for small $\Delta t$. To zeroth order for the
momenta and order $\Delta t $ for the fields we obtain
\begin{eqnarray}
P[\{\pi,\phi\}] = \exp\Big(- \Delta x \sum_i \big[ &&\beta' 
{\pi_i^2 \over 2} + \beta S(\phi_i)  \label{e7}\\&&
- \Delta t  {\beta \over
2} \pi_i  {\partial S \over \partial \phi_i} \big]\Big), 
\nonumber
\end{eqnarray}
where $\beta'=\beta \left( 1+ \Delta t {\eta \over 2}\right)$.  Note
that the discretization induces cross terms between $\phi$ and $\pi$
in the canonical distribution. This is a general feature of higher
order solutions in $\Delta t $. (These terms rapidly become very
complicated.)  Different time discretizations lead to different
discrete time Fokker-Planck equations.  The numerical simulations
described below employed a stochastic second order Runge-Kutta
algorithm \cite{num}.

The equilibrium density of configurations
of the space- and time-discretized theory, is obtained by
performing the Gaussian integral over the momenta in \eqref{e7}:
\begin{eqnarray}
P[\{\phi\}] &&= A\exp\left[- \beta \Delta x { \sum_i}
 \left\{ S(\phi_i) - {\Delta t^2  \over 8}
\left( {\partial S \over \partial \phi_i} \right)^2 \right\} \right]. 
\label{e8}
\end{eqnarray}
The effect of the time discretization is explicitly seen as a
modification of the equilibrium density. Further integration of
\eqref{e7} cannot be performed so easily because each \(S(\phi_i)\)
depends also on \(\phi_{i+1}\).

The functional integral of $P[\phi]$ over $\phi$ defines the
configurational partition function
\begin{equation}
Z_{\phi}= Z_\pi \int D\phi\hbox{e}^{-\beta S[\phi]}=\prod_{i=1}^N\int
\dd\bar{\phi}_i\hbox{e}^{-\beta \Delta S[\phi_{i+1},\phi_i]}.
\end{equation}
which we study in the next section. Here $\dd\bar{\phi}_i = \bar{N} 
\dd \phi_i$, with $\bar{N} = \sqrt{\beta \over 2 \pi \Delta x}$.

In principle, the partition function as calculated above would include
artifacts from both the time and space discretizations. In actual
computational practice, given that we can estimate the order of the
time stepping errors as shown already, it is not difficult to reduce
the time step to a level where the remaining errors are suppressed
compared to the errors from the spatial discretization. Once this is
done, we may safely ignore the discretization in time and concentrate
solely on the errors due to the spatial lattice.

\section{The Transfer Integral}

To compute the partition function explicitly we make use of the
transfer integral method \cite{ti}.  The configurational partition
function $Z_{\phi}$ is given by
\begin{eqnarray}
Z_{\phi}&=&
\int_{-\infty}^{\infty} \dd{\bar \phi}_1\ldots\dd{\bar \phi}_N
\prod_{i=1}^N
T(\phi_{i},\phi_{i+1}),
\label{Zphi}
\end{eqnarray}
where 
\begin{eqnarray}
&&T(\phi_{i},\phi_{i+1})= \nonumber \\
&&\exp\left\{-\half\beta\Delta x\left[
\left({\phi_{i+1}-\phi_i\over\Delta x}\right)^2
+ V(\phi_{i}) + V(\phi_{i+1}) \right]\right\}
\nonumber
\end{eqnarray}
and\(\quad
\phi_{N+1} = \phi_1\) implements spatially periodic boundary
conditions. The difficulty with evaluating $Z_{\phi}$ lies in the
coupling of integrals at different space points.  The idea behind the
transfer operator method is to ``localize'' the evaluation of the
integrals in \eqref{Zphi}.

The {transfer operator} \(\hat T\) is defined as follows
\begin{equation}
\hat T\psi(\phi_{i+1}) = 
\int_{-\infty}^{\infty}\dd{\bar\phi}_iT(\phi_{i},\phi_{i+1})\psi(\phi_{i}).
\end{equation}
Suppose we can find the eigenvalues of \(\hat T\). That is, suppose
we can solve the following Fredholm equation:
\begin{equation}
\int_{-\infty}^{\infty} \dd{\bar \phi}_i T(\phi_{i},\phi_{i+1})
\psi_n({\phi}_i) =
t_n\psi_n({\phi}_{i+1})~,
\label{transop}
\end{equation}
where the \(t_n\) are positive constants that we write
for later convenience as
\begin{equation}
t_n=\hbox{e}^{-\beta\Delta x\epsilon_n}.
\label{endef}
\end{equation}
Then
\begin{equation}
Z_{\phi}= \sum_nt_n^N.
\label{zeval}
\end{equation}
In the limit \(N\to\infty\), the sum \eqref{zeval} is dominated by
the largest eigenvalue:
\begin{equation}
Z_{\phi}=\sum_nt_n^N \to t_0^N=\hbox{e}^{-\beta L\epsilon_0},
\end{equation}
where $L=N\Delta x$ is the physical length of the lattice. In the
thermodynamic limit $L\to \infty$, the free energy density is simply
$F_{\phi}=\epsilon_0$. It is clear that once the partition function is
known in the thermodynamic limit we may compute from it any
thermodynamic quantity. Moreover, it is possible to show that spatial,
and in linear response theory, temporal correlation functions can also
be computed via a knowledge of the spectrum of the transfer operator
\cite{tempcorr}.

We now turn to the procedure for solution of \eqref{transop} by first
converting it into an infinite order partial differential equation.
We first rewrite (\ref{transop}) as
\begin{eqnarray}
&&\hbox{e}^{-\half\beta\Delta x V(\phi_{i+1})}\times\nonumber\\
&&\int \dd{\bar \phi}_i~\hbox{e}^{-{\beta\over2\Delta x}
(\phi_{i+1}-\phi_i)^2}\hbox{e}^{(\phi_{i}-\phi_{i+1}) 
\partial/\partial\phi_{i+1}}
\chi(\phi_{i+1})
\nonumber\\
&&=\hbox{e}^{-\beta\Delta x\epsilon_n}\psi_n({\phi}_{i+1})~,
\end{eqnarray}
where
\begin{equation}
\chi(\phi) = 
\hbox{e}^{-\half\beta\Delta xV(\phi)}\psi_n({\phi}).
\label{chidef}
\end{equation}
The special form of the Fredholm kernel has led to a simple Gaussian
integral that yields:
\begin{eqnarray}
&&\hbox{e}^{-\frac12\beta\Delta x V(\phi_{i+1})} 
\hbox{e}^{(\Delta x/2\beta){\partial^2/\partial\phi_{i+1}^2}}
\left(
\hbox{e}^{-\frac12\beta\Delta x V(\phi_{i+1})} 
\psi_n({\phi}_{i+1})
\right)\nonumber\\
&&=\hbox{e}^{-\beta\Delta x\epsilon_n}\psi_n({\phi}_{i+1})~.
\label{result}
\end{eqnarray}
This (exact) result yields the form
$\hbox{e}^U\hbox{e}^D\hbox{e}^U\psi =\hbox{e}^C\psi$ where $U$ and $D$
are operators and $C$ is a real number. The Campbell-Baker-Hausdorff
series in this case is formally an expansion in powers of $\Delta x$.
To linear order in $\Delta x$, the CBH expansion applied to
\eqref{result} yields:
\begin{equation}
\hbox{e}^{-\beta\Delta x V(\phi)+{\Delta x\over
2\beta}{\partial^2\over\partial\phi^2}}\psi_n({\phi})=
\hbox{e}^{-\beta\Delta x\epsilon_n}\psi_n({\phi})~,
\label{sec}
\end{equation}
or equivalently
\begin{equation}
\left[-{1\over 2\beta^2}{\partial^2\over\partial\phi^2}+V(\phi)\right]
\psi_n=\epsilon_n\psi_n~.
\label{sec2}
\end{equation}
The transfer integral technique thus reduces the calculation of
\(Z_{\phi}\) to the calculation of the eigenvalues $\epsilon_n$ of a 
corresponding Schr\"odinger equation:
\begin{eqnarray}
 \left\{-{1\over2\beta^2}{\partial^2\over\partial\phi^2}
+U(\phi,\Delta x)\right\}\psi_n=\epsilon_n\psi_n~, 
\label{se}
\end{eqnarray}
where \(U(\phi,0)=V(\phi)\).  The calculation is explicitly performed
on the lattice, at finite $\Delta x$: leading order corrections to the
eigenvalues of the Schr\"odinger equation \eqref{se} are proportional
to \(\Delta x^2\). For the problem at hand, one finds \cite{TS},
\begin{eqnarray}
&& \left\{-{1\over
2\beta^2}{\partial^2\over\partial\phi^2}+V(\phi)+{1\over 6}(\Delta
x)^2 \left [{1\over 4}\left({\partial V \over \partial\phi}\right)^2 
\right.\right. 
\label{schrocorr1} \\
&& \left.\left. + {1\over 2\beta^2}{\partial^2 V \over\partial\phi^2}
{\partial^2\over\partial\phi^2} 
+{1\over 2\beta^2}{\partial^3 V \over\partial\phi^3}
{\partial\over\partial\phi} 
+{1\over 8\beta^2} {\partial^4 V \over\partial\phi^4} \right]
\right\}\psi_n=
\epsilon_n\psi_n~. \nonumber 
\end{eqnarray}
Higher order corrections in $\Delta x$ in \eqref{schrocorr1} can be
computed in a tedious though straightforward fashion by going to
higher orders in the CBH expansion. It is easy to show from the
symmetric form of (\ref{result}) and the Hermitian/anti-Hermitian
alternation of terms in the CBH expansion that the error terms are
always of even order in powers of $\Delta x$. Thus, if a method is 
found to cancel errors up to a certain order $m$, it 
automatically reduces the error to order $m+2$.

The simplest example of \(\Delta x\) dependence is the free theory:
 $V={1 \over 2} \phi^2$. Then \eqref{schrocorr1} reduces to
\begin{eqnarray}
&& \left[ -{1\over 2 {\beta'}^2}{\partial^2\over\partial\phi^2}
+ {1 \over 2} {m'}^2 \phi^2 \right] \psi_n = 
\epsilon_n\psi_n,
\end{eqnarray}
with $\beta'= \beta/\sqrt{1-a}$, $m'= \sqrt{1 +a/2}$ and $a= (\Delta
x)^2/6$.  This implies in particular for the energy spectrum
$\epsilon_n = \left(n+{1\over 2}\right){m' \over \beta'}\simeq
(n+{1\over 2}){1 \over \beta} \left(1 - {1\over 24}(\Delta
x)^2\right)$.  The free theory is a convenient special case because
the corresponding SPDE is linear and exactly solvable. Quantities such
as $\mean{\phi^2(x)}$ can be evaluated exactly and compared to the
results from the transfer integral. Both procedures agree, and
\begin{equation}
\mean{\phi^2(x)}= \frac{1}{2\beta} { 1 \over \sqrt{1 + 
{\Delta x^2 \over 4}}} 
= \frac{1}{2\beta}\left(1- \frac18\Delta x^2\right) + O(\Delta x^4). 
\label{freeps}
\end{equation}
Note that the leading dependence on lattice spacing is proportional to
\(\Delta x^2\).

We now turn to the question of lattice errors in determining the kink
density, which, at sufficiently low temperatures, is controlled
completely by the correlation length derived from the two-point
function $\langle\phi(0)\phi(x)\rangle$. Applying the transfer
integral formalism, it is easy to show that this correlation function
is a sum of exponentials with exponents proportional to differences of
eigenvalues of \eqref{se}. The correlation length is determined by the
energy difference between the ground and first excited states of
\eqref{se} \cite{ti}. At low temperatures, the WKB (or semiclassical)
approximation is excellent and this energy difference is the
exponentially small tunnel-splitting term.  Note that at low
temperatures the kink density is given directly by the correlation
length, $\rho_k \simeq 1/(4\lambda_{\infty})$ \cite{fash}.

At low temperatures the first two eigenfunctions of \eqref{sec2} are of
the form 
\begin{eqnarray}
\psi_S={1\over\sqrt{2}}(\psi_L+\psi_R), \nonumber\\
\psi_A={1\over\sqrt{2}}(\psi_L-\psi_R),   \label{two psi}
\end{eqnarray}
where $\psi_S$ is the (symmetric) ground state and $\psi_A$ is the
(antisymmetric) first excited state. Here $\psi_L$ and $\psi_R$ are
the usual localized states, one on each side of the barrier.
To estimate the error due to finite lattice spacing we use first order
perturbation theory in $(\Delta x)^2$. The corrected energies are
then,
\begin{eqnarray}
E_0^{(\Delta x)} = E_0+\langle\psi_S|\delta H|\psi_S\rangle, \nonumber\\
E_1^{(\Delta x)} = E_1+\langle\psi_A|\delta H|\psi_A\rangle,
\end{eqnarray}
where $E_0$ and $E_1$ are the results from the continuum theory and
$\delta H \sim O(\Delta x^2)$ is the error Hamiltonian.  It follows
that the energy difference is
\begin{equation}
\Delta E_{10}^{(\Delta x)}=\Delta E_{10}-2\langle\psi_L|\delta
H|\psi_R\rangle. 
\end{equation}
The error Hamiltonian can be read off from \eqref{schrocorr1} and it
is clear that the error in energy differences, and hence kink density
at low temperatures, is also $O(\Delta x^2)$ at leading order.
Corrections to the the eigenstates lead to higher order \(\Delta x \)
dependences.

More generally, given any eigenvector $|\psi\rangle$ of the continuum
Schr\"odinger equation, for the specific form of $\delta H$ of
\eqref{schrocorr1}, integration by parts and use of \eqref{se} yields
\cite{TS},
\begin{eqnarray}
\langle\psi|\delta H|\psi\rangle = -{(\Delta x)^2 \over 24} 
\langle\psi| \left( {\delta V \over \delta \phi}\right)^2
|\psi\rangle.  
\label{change}
\end{eqnarray}
Apart from the eigenvectors, there is no temperature dependence in
\eqref{change}. This remarkable fact immediately suggests the
introduction of a counterterm in the lattice equations which, in
perturbation theory, would lead to the cancellation of errors of order
$(\Delta x)^2$. Modifying the potential as follows
\begin{equation}
V(\phi) = V(\phi) 
- {(\Delta x)^2 \over 24} \left( {\delta V \over \delta
\phi}\right)^2, 
\label{cterm}
\end{equation}
leads to the cancellation of lattice dependences to order $(\Delta
x)^2$ in a way that preserves the fluctuation-dissipation relation
(taken at any temperature) and is thus suited for dynamics as well as
thermodynamics. With $\Delta x$ taken to be small enough, the leading
error now becomes dominantly $O(\Delta x)^4$. We note that unlike the
situation for PDEs where one improves the lattice approximation for
spatial derivatives, here a local counterterm produces the same
effect. 

In the specific case of a $\phi^4$ potential, the counterterm
\eqref{cterm} gives a new potential including the term $-\Delta x^2
\phi^6/24$. The corrected potential is thus unbounded from below! In
first order perturbation theory this is not a problem since the
corresponding wave function is exponentially small in the pathological
region of the compensated potential \cite{TS}. However, if the full
potential is to be used in a Langevin simulation it is clear that at
sufficiently long times, the unboundedness of the potential implies
the nonexistence of a stable thermal distribution. Fortunately, it is
simple to estimate whether this problem actually shows up in real
simulations. The answer, as we show below, is that it is of absolutely
no practical significance in the parameter range of interest.

The resolution of this apparent difficulty brings us back to the
validity of the expansion in $\Delta x$.  So far we have implicitly
assumed that $\beta \simeq 1$, so that $\Delta x$ is the only small
parameter and controls the order of the expansion. If on the other
hand one wanted to work in a regime where $\Delta x \geq \beta$, the
whole expansion in $\Delta x$ would have to be rederived in terms of
an appropriate small parameter. In any case this latter regime would
always constitute a poor approximation to the continuum: It is the
Ising (disorder) limit of the field theory.

A simple argument for why the counterterm works at small temperatures,
meaning $\Delta x \ll \beta$, is the following. Consider a temperature
large relative to the potential barrier between the minima. Then, from
the uncorrected eigenvalue equation, $\langle \phi(x)^2 \rangle \simeq
\beta^{-1}$. On the other hand the value of $\phi(x)^2$ for which a
fluctuation can probe the effect of the negative $\phi^6$ term at
large $\phi$ is $\phi^2(x) \simeq 6/(\Delta x)^2$.  Therefore the
condition for the negative $\phi^6$ term not to affect the evolution
is $\Delta^2 x \ll 6 \beta$. At lower temperatures it is more
appropriate to explicitly calculate the Kramers escape rate
\cite{risken} across the barrier separating the metastable and
unstable regions of the compensated potential. Assuming the lattice
sites to be uncoupled (this gives an overestimate of the true rate),
the calculation yields $\Gamma_K\sim\exp(-4\beta/3(\Delta x)^4)$,
which turns out to be vanishingly small in practice: For $\Delta
x=.5$, $\beta=5$, and a lattice size of $10^6$ points, the probability
of an escape at a single site per unit time is only $\sim
10^{-41}$. In our numerical calculations we have verified that the
counterterm can indeed be successfully used in the appropriate
circumstances with no hint of any instabilities.

\section{The One Loop Approach}

In contrast to the above considerations, the 1-loop counterterm
proposed in Ref. \cite{GM} arises from the conjecture that the leading
dependence of the partition function on $\Delta x$ coincides with the
most divergent term for the same theory in higher dimensions. Although
the relevant computations are well known we will spell out some of the
steps to make every assumption clear. The basic idea is to start again
with the canonical partition function:
\begin{equation}
Z = N \int D\phi e^{- \beta S[\phi]}~.
\end{equation}
The field $\phi$ is then decomposed into a
background field $\phi_b$ and a fluctuation field $\chi$, $\phi =
\phi_b + \chi$, and assuming the fluctuations to be small, expanded
around \(\phi_b\):  
\begin{equation}
S[\phi_b + \chi] \approx S[\phi_b] + {\delta S \over \delta 
\phi }_{ \vert_{\chi=0}}
\chi + {1 \over 2} \chi {\delta^2 S \over \delta \phi^2 
}_{\vert_{\chi =0}} \chi + \ldots.
\end{equation}

If $\phi_b$ is an extremum of $S[\phi]$ then the first term 
vanishes. Under this assumption
\begin{equation}
Z = N e^{-\beta S[\phi_b]}  \int D\chi e^{- \beta {1 \over 2} 
\chi {\delta^2 S \over \delta \phi^2}_{ \vert_{\chi=0}} \chi}.
\end{equation}
Because ${\delta^2 S \over \delta \phi^2 }_{\vert_{\chi=0}}$ is
independent of $\chi$ the functional integration is strictly Gaussian
and can be performed exactly: 
\begin{equation}
 N \int D\chi e^{- \beta {1 \over 2} 
\chi {\delta^2 S \over \delta \phi^2} \vert_{\chi=0} \chi} =
{ \rm Det}^{-1/2} \left( S_I^{''} \over S_0^{''} \right).
\end{equation}
Here we have adopted the usual normalization to the free theory. The
action $S = S_0 + S_I$, was decomposed into the action for the free
theory $S_0$ (gradient and mass terms) and the interactions $S_I$.
Primes denote functional derivatives relative to $\phi$.

This can be written as
\begin{eqnarray}  
{\rm Det}^{-1/2} \left( S^{''} \over S_0^{''} \right) &=&
{\rm Det}^{-1/2} \left( 1 + K \right) \nonumber\\
&=& e^{- {1\over 2} {\rm Tr}\log(1 + K)}, 
\end{eqnarray}
where $K = S_I^{''}/S_0^{''}$. Performing the 1D $k$-space trace
integral, ($m=0$), with an ultraviolet cutoff $\Lambda = \pi/{\Delta
x}$, we obtain one loop corrections to the potential
\begin{equation}
V_{1L}(\phi,\Lambda) = V_0 + {T \over 4} \sqrt{S_I^{''}(\phi_b)} 
- {T \Delta x \over 4 \pi^2} S_I^{''}(\phi_b)~.
\label{gm1}
\end{equation}
The partition function is now approximately given by
\begin{equation}
Z =  e^{-\beta S[\phi_b] + {1 \over 4 \beta} \sqrt{S_I^{''}(\phi_b)} 
- {\Delta x \over 4 \beta  \pi^2} S_I^{''}(\phi_b) }~.
\label{gm2}
\end{equation}

Equations (\ref{gm1}) and (\ref{gm2}) constitute the basis for the 
proposal of Ref. \cite{GM}. In order to cancel the leading 
$\Delta x$ dependence arising in this scheme, the original 
bare potential is modified by the addition of the last  
term in Eq.~(\ref{gm1}), (with a positive sign).  

Notice that while a careful accounting of the dynamics on the lattice
yields a leading correction of order $(\Delta x)^2$, regardless of any
assumptions about the dominant thermodynamic field configurations, the
one loop procedure leads to a correction of order $\Delta x$. In
contrast to the correct answer discussed in the previous section, the
one loop procedure gives no corrections for the free theory since in
this case $S_I\equiv 0$.

\section{Comparison with Numerical Solutions}

Accurate Langevin studies of even one-dimensional field theories
require large lattices and long running times. It has only recently
been realized, by comparison against exact analytic results for
nonlinear field theories, that fairly large errors (e.g., $30\%$ or
greater in the kink density) can easily arise if numerical studies are
not carried out with careful error control methodologies \cite{qes}.

In order to test the predictions of the previous sections, we ran
large scale Langevin evolutions with typical lattice sizes \(N=10^6\),
and with the time step related to the lattice spacing by \(\Delta t =
0.05\Delta x^2\).

\begin{figure}
\vspace{0.1cm} 
\epsfig{file=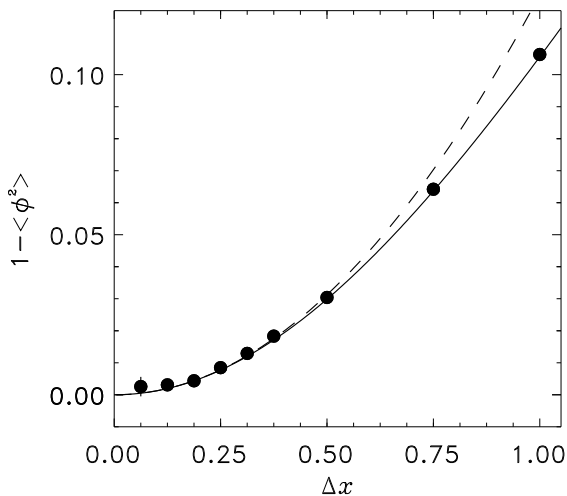,width=3.2in, clip=}
\vspace{.3cm}
\caption{The dependence of $1-\langle \phi^2 \rangle$ on $\Delta x$,
for $V(\phi)=\phi^2/2$. The numerical results ($\bullet$) at $\beta=2$
are compared with the exact equilibrium result \eqref{freeps} (solid
line). The dashed line shows the Taylor expansion of \eqref{freeps} to
order $\Delta x^2$. Statistical error bars are not shown if they
are smaller than the symbol size.}
\label{fig1}
\end{figure}

A first test which allows comparison against exact analytical results
are the lattice dependences for the linear SPDE (free theory) defined
by $V(\phi)=\phi^2/2$. Figure~\ref{fig1} shows the 
plot of the thermal equilibrium
$1-\mean{\phi^2}$ versus $\Delta x$. The numerical data are in
excellent agreement with the (exact) theoretical predictions.

In the more general case of a nonlinear SPDE, we cannot expect
explicit exact solutions for arbitrary $\Delta x$, but thermodynamic
quantities can be obtained to order $\Delta x^2$ from the eigenvalues
of the perturbed Schr\"odinger equation extracted from the transfer
integral, as described in Section III. In the case of predictions for
the kink density, precise comparison with numerical results has not
been possible until recently, partly due to the difficulty of counting
the number of kinks in a noisy configuration. The correlation length
is, however, a well-defined quantity at any temperature, independent
of kink-counting schemes. 

We extract the correlation length \(\lambda_{\infty}\) from the
numerically determined field configurations as follows. Let
\begin{equation}
c(i\Delta x) = {\mean{\phi(j)\phi(j+i)}},
\label{cdef}
\end{equation}
and
\begin{equation}
\lambda(x) = \Delta x 
\left(\log\left(\frac{c(x)}{c(x+\Delta x)}\right)\right)^{-1}.
\label{ldef}
\end{equation}
The correlation length is \(\lim_{x\to\infty}\lambda(x)\):
\begin{equation}
\langle\phi(0)\phi(x)\rangle\to\exp\left(-x/\lambda_{\infty}\right),
\qquad x\to\infty.
\end{equation}

The correlation function \(c(x)\) is in general a sum of exponentials
(the smallest exponent being the correlation length). For values of
\(x\) much smaller than the correlation length, therefore,
\(\lambda(x) < \lambda_{\infty}\). In practice, for large \(x\),
finite statistics mean that the ratio in \eqref{ldef} cannot be
evaluated precisely. One therefore evaluates the correlation length by
plotting \(\lambda(x)\) versus \(x\) and looking for a plateau at
intermediate values of \(x\).

\begin{figure}
\vspace{0.1cm} 
\epsfig{file=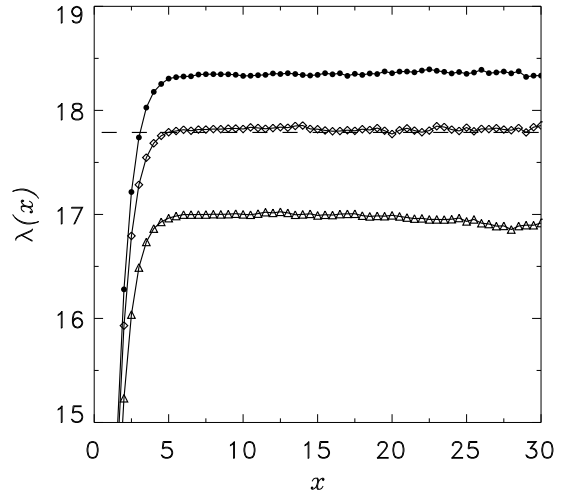,width=3.2in, clip=}
\vspace{.3cm}
\caption{ The correlation length $\lambda$ computed with the bare
potential ($\bullet$), the counterterm of \eqref{cterm} ($\diamond$)
and the one loop counterterm ($\triangle$), for $\Delta x=0.5$ and
$\beta=5$. The dashed line shows the continuum exact result, computed
via the transfer integral. }
\label{counter}
\end{figure}

We measured the correlation length using three different Langevin
evolutions: (a) A standard simulation using a second-order stochastic
Runge-Kutta integrator; (b) A simulation with the counterterm
\eqref{cterm}; (c) A simulation with the counterterm proposed in
Ref. \cite{GM}. Results for $\Delta x=0.5$ are shown in Figure
\ref{counter}. The counterterm \eqref{cterm} shifts the result from
the Langevin evolutions (a) on the lattice very close to the exact
continuum result, shown as a dashed line. The standard simulation
overestimates $\lambda_{\infty}$, whereas the one loop counterterm
results in an underestimate with an error larger than the ``bare''
simulation (a) without any counterterm.

\begin{figure}
\vspace{0.1cm} 
\epsfig{file=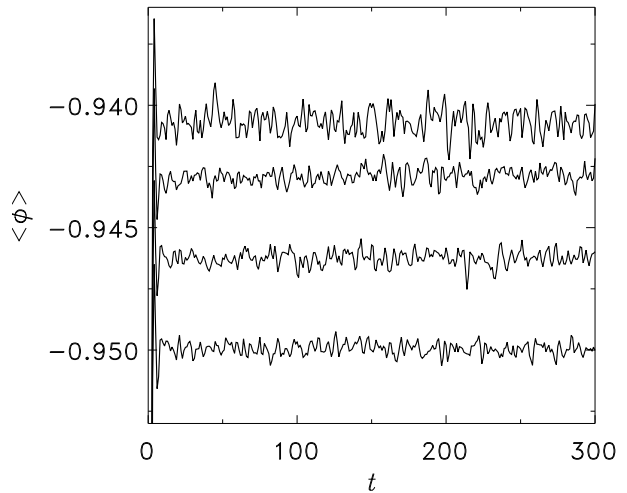,width=3.2in, clip=}
\vspace{.3cm}
\caption{ Early evolution of the space-averaged mean value of
\protect\(\phi\) for different values of the lattice spacing \(\Delta
x\). From top to bottom, the lattice spacings are: $\Delta
x=0.25,~0.5,~0.75,~1.0.$ We used lattices of 1048576 points and \(\Delta t
= 0.05\Delta x^2\). \protect\(\beta=10\), \protect\(\eta=1\).}
\label{gmdx}
\end{figure}

\begin{figure}
\vspace{0.1cm}\epsfig{file=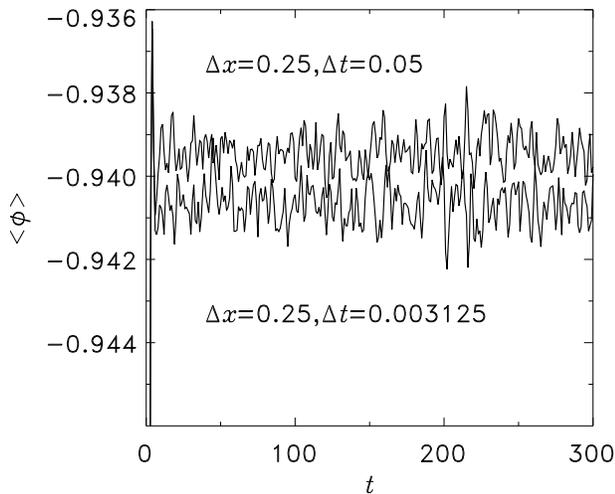,width=3.2in, clip=}
\vspace{.3cm}
\caption{ Early evolution of the space-averaged mean value of
\protect\(\phi\) for two values of the time step \(\Delta
t\), with \protect\(\beta=10\), \protect\(\eta=1\).  }
\label{gmdt}
\end{figure}

As a further test we repeated, with large lattices and smaller
time steps, a numerical experiment presented in Ref. \cite{GM}. The
initial condition is chosen uniform at the minimum of \(V(\phi)\),
\(\phi_0=-1\); the system is then run for a short time (before any
kinks appear) so as to observe the relaxation to a mean value
\(\phi_m\).  Although this does not result in a strictly thermalized
configuration, small wave-length fluctuations quickly display a thermal
spectrum. (In other words, the ``phonon'' relaxation time is much smaller
than the timescale for kink nucleation.)  In Figure \ref{gmdx} we show
the value of \(\mean{\phi}\) as a function of time for four values of
\(\Delta x\).  From the plateau for moderate times, we can obtain a
fairly precise estimate of \(\phi_m\). As a cautionary note, we point
out that at small \(\Delta x\), a small stepsize is also needed (see
Figure \ref{gmdt}).

\begin{figure}
\vspace{0.1cm} 
\epsfig{file=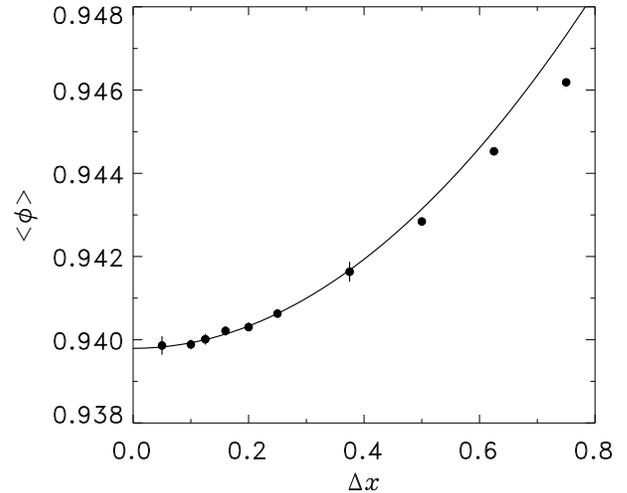,width=3.2in, clip=}
\vspace{.3cm}
\caption{ Space- and time-averaged mean value of \protect\(\phi\) for
different values of the lattice spacing \(\Delta x\).  The solid line
is the large-\(\beta\) estimate \eqref{phi0pred} obtained from the
Gaussian ansatz \eqref{gauss} with \protect\(\beta=10\),
\protect\(\eta=1\). Statistical error bars are not shown if they are
smaller than the symbol size.}
\label{gmdxav}
\end{figure}

It is possible to employ a Gaussian approximation (following
Ref. \cite {fash}) to obtain a rather good estimate of \(\phi_m\) as a
function of \(\Delta x\), the result being shown in Figure
\ref{gmdxav}: The leading dependence, both analytically and
numerically, is clearly quadratic in $\Delta x$. To obtain the
analytic result, we use the fact that the probability density of
\(\phi\) is the square of the ground state of the Schr\"odinger
equation \eqref{se}. (This density emerges from dynamic
simulations or calculations; it is not an input to numerics.) 

We proceed further by using a Gaussian ansatz \cite{fash} for the
ground state eigenfunction:
\begin{equation}
\psi_0(\phi) = \left(\frac{\Omega}{\pi}\right)^{\frac12}
\exp(-\frac12\Omega(\phi-\phi_0)^2).
\label{gauss}
\end{equation}
The parameters \(\Omega\) and \(\phi_0\) are
obtained by minimizing the energy
\begin{equation}
E_0 = \int_{-\infty}^{\infty}\psi_0^2(\phi)H(\phi)\dd \phi.
\label{gausse}
\end{equation}
For large \(\beta\) the two free parameters are related by
\begin{equation}
\Omega = \beta\left(3\phi_0^2 - 1 \right)^{\frac12} + {\cal O}(1),
\label{omphi0}
\end{equation}
and 
\begin{equation}
E_0 = -\frac12\phi_0^2 + \frac14\phi_0^4 
+ \beta^{-1}\frac1{2}(3\phi_0^2-1)^{\frac12} +{\cal O}(\beta^{-2}).
\label{gausseapp}
\end{equation}

The dependence of \(\phi_m\) on \(\Delta x\) can now be estimated using 
the Gaussian approximation of \eqref{gauss}. At finite \(\Delta x\),
we replace \(V(\phi)\) in \eqref{gausse} by \(V(\phi) + 
\frac1{24}\Delta x^2 (\frac{\delta}{\delta \phi}V(\phi))^2\).
Minimizing \eqref{gausseapp} with respect to \(\phi_0\) gives
\begin{equation}
\phi_0(\Delta x) = \phi_0(0)  + \beta^{-1}\Delta x^2\frac{11}{64\sqrt{2}}
+ {\cal O}(\beta^{-2})~,
\label{phi0pred}
\end{equation}
which is plotted in Figure \ref{gmdxav}, in excellent agreement with
the numerical results.

\section{Conclusions}

We have presented a complete procedure to identify space and time
discreteness effects in Langevin studies of 1+1-dimensional field
theories on the lattice.  This scheme permits the determination of the
correct continuum limit of the theory in thermal equilibrium.  In
particular, we have shown that for the standard spatial discretization
of the Langevin equation, quantities of interest such as the kink
density and the expectation value of the field and its variance differ
from the continuum values by terms of order $\Delta x^2$.  High
resolution numerical results are in excellent agreement with our
analytical predictions.
  
In any numerical Langevin evolution errors result from the necessary
discretization of a field theory in both time and space.  The effect
of the former is to modify the form of the canonical distribution as
seen from the solution of the corresponding Fokker-Planck
equation. The use of higher order timestepping algorithms can render
this error subdominant when compared to errors arising from the
discretization of the spatial lattice.

This spatial discretization error can be computed systematically in
powers of $(\Delta x)^2$ via the use of the transfer integral to solve
for the partition function on the lattice.  This procedure leads to
the identification of a simple local counterterm which in turn permits
the practical elimination of the leading order lattice error in
Langevin evolutions at low temperature.

For the \(\phi^4\) theory in one space dimension, the density of kinks
converges to a well-defined value at any temperature low enough that
kinks are clearly separated from small wave-length fluctuations (or
``phonons''). In practice this is essentially the range of
temperatures where the dilute gas approximation (which is equivalent
to a WKB solution of the transfer integral) is valid.  Precision
calculations over a wide range of temperatures that agree with
transfer integral predictions are reported in Ref. \cite{qes}.  For
quantities that are defined unambiguously at arbitrary temperatures,
such as the correlation length, results based on the WKB approximation
to the transfer integral will fail at sufficiently high
temperatures. This is independent of lattice errors and does not
preclude analytical and numerical study using lattice simulations.

Our results disagree with those reported by Gleiser and M\"uller based
on a one loop counterterm \cite{GM}. A critical examination of their
proposal has shown that it does not in fact constitute a scheme for
the control and elimination of lattice errors.  We have also carried
out a direct comparison of our numerical results with those presented
in Ref. \cite{GM} and are led to conclude that their data must have
been of insufficient quality to quantitatively characterize lattice
spacing dependences.

Finally we wish to note that our methods strongly depend on the use of
the transfer integral to solve (exactly) for the non-perturbative
field thermodynamics. The application of the procedure described above
is in general guaranteed for any (local) field theory in one spatial
dimension, requiring only the choice of the appropriate potential.

In higher dimensions such a solution becomes increasingly
difficult. Nevertheless, for the particular case of two dimensions
several lattice models can be solved exactly, precisely by applying
the transfer integral technique \cite{Baxter}.  The eigenvalue
problem, elegantly posed in terms of an ordinary time-independent
Sch\"odinger equation in one dimension, now amounts to solving for the
eigenvalues of an infinite  matrix.  Regardless of this
apparent difficulty, exact solutions are known in several interesting
cases, most notably perhaps for the 2D Ising model on a square
lattice. It is therefore conceivable that the detailed approach to the
continuum in these models can be understood via the procedure
described above.

\section{Acknowledgements}

We wish to acknowledge very helpful contributions from Nuno Antunes
who participated in the early part of this work. Large scale
computations were performed on the T3E at the National Energy Research
Scientific Computing Center (NERSC) at the Lawrence Berkeley National
Laboratory.

\end{document}